# Integrated one- and two-photon scanned oblique plane illumination (SOPi) microscopy for rapid volumetric imaging

**MANISH KUMAR, SANDEEP KISHORE, JORDAN NASENBENY, DAVID L. MCLEAN, AND YEVGENIA KOZOROVITSKIY**[*]

*Department of Neurobiology, Northwestern University, Evanston, IL 60208, USA*
*\*yevgenia.kozorovitskiy@northwestern.edu*

**Abstract:** Versatile, sterically accessible imaging systems capable of *in vivo* rapid volumetric functional and structural imaging deep in the brain continue to be a limiting factor in neuroscience research. Towards overcoming this obstacle, we present integrated one- and two-photon scanned oblique plane illumination (SOPi, /sōpī/) microscopy which uses a single front-facing microscope objective to provide light-sheet scanning based rapid volumetric imaging capability at subcellular resolution. Our planar scan-mirror based optimized light-sheet architecture allows for non-distorted scanning of volume samples, simplifying accurate reconstruction of the imaged volume. Integration of both one-photon (1P) and two-photon (2P) light-sheet microscopy in the same system allows for easy selection between rapid volumetric imaging and higher resolution imaging in scattering media. Using SOPi, we demonstrate deep, large volume imaging capability inside scattering mouse brain sections and rapid imaging speeds up to 10 volumes per second in zebrafish larvae expressing genetically encoded fluorescent proteins GFP or GCaMP6s. SOPi's flexibility and steric access makes it adaptable for numerous imaging applications and broadly compatible with orthogonal techniques for actuating or interrogating neuronal structure and activity.

**OCIS codes:** (180.2520) Fluorescence microscopy; (180.6900) Three-dimensional microscopy.

## References and links

1.  D. A. Dombeck, C. D. Harvey, L. Tian, L. L. Looger, and D. W. Tank, "Functional imaging of hippocampal place cells at cellular resolution during virtual navigation," Nat. Neurosci. **13**(11)**,** 1433-1440 (2010).
2.  S. W. Hell, and J. Wichmann, "Breaking the diffraction resolution limit by stimulated emission: stimulated-emission-depletion fluorescence microscopy," Opt. Lett. **19**(11), 780-782 (1994).
3.  M. J. Rust, M. Bates, and X. W. Zhuang, "Sub-diffraction-limit imaging by stochastic optical reconstruction microscopy (STORM)," Nature Methods **3**(10), 793–795 (2006).
4.  S. T. Hess, T. P. K. Girirajan, and M. D. Mason, "Ultra-high resolution imaging by fluorescence photoactivation localization microscopy," Biophys. J. **91**(11), 4258–4272 (2006).
5.  M. G. L. Gustafsson, "Surpassing the lateral resolution limit by a factor of two using structured illumination microscopy," J. Microsc. **198**(2), 82–87 (2000).
6.  P. Theer, M. T. Hasan, and W. Denk, "Two-photon imaging to a depth of 1000 μm in living brains by use of a Ti: $Al_2O_3$ regenerative amplifier," Opt. Lett. **28**(12), 1022-1024 (2003).
7.  M. Levoy, R. Ng, A. Adams, M. Footer, and M. Horowitz, "Light field microscopy," ACM Trans. Graph. **25**(3), 924 (2006).
8.  T. Nöbauer, O. Skocek, A. J. Pernía-Andrade, L. Weilguny, F. M. Traub, M. I. Molodtsov, and A. Vaziri, "Video rate volumetric $Ca^{2+}$ imaging across cortex using seeded iterative demixing (SID) microscopy," Nature Methods **14**(8), 811–818 (2017).
9.  H. Siedentopf, and R. Zsigmondy, "Uber Sichtbarmachung und Größenbestimmung ultramikoskopischer Teilchen, mit besonderer Anwendung auf Goldrubingläser," Ann. Phys. **315**(1), 1–39 (1902).
10. J. Huisken, J. Swoger, F. D. Bene, J. Wittbrodt, and E. H. K. Stelzer, "Optical sectioning deep inside live embryos by selective-plane illumination microscopy," Science **305**(5686), 1007–1009 (2004).
11. C. J. Engelbrecht, and E. H. Stelzer, "Resolution enhancement in a light-sheet-based microscope (SPIM)," Opt. Lett. **31**(10), 1477-1479 (2006).
12. P. J. Keller, A. D. Schmidt, J. Wittbrodt, and E. H. K. Stelzer, "Reconstruction of zebrafish early embryonic development by scanned light-sheet microscopy," Science **322**(5904), 1065–1069 (2008).

13. N. Vladimirov, Y. Mu, T. Kawashima, D. V. Bennett, C. T. Yang, L. L. Looger, P. J. Keller, J. Freeman, and M. B. Ahrens, "Light-sheet functional imaging in fictively behaving zebrafish," Nature Methods **11**(9), 883 (2014).
14. P.G. Pitrone, J. Schindelin, L. Stuyvenberg, S. Preibisch, M. Weber, K. W. Eliceiri, J. Huisken, and P. Tomancak, "OpenSPIM: an open-access light-sheet microscopy platform," Nature Methods **10**(7), 598–599 (2013).
15. Y. Wu, A. Ghitani, R. Christensen, A. Santella, Z. Du, G. Rondeau, Z. Bao, D. Colón-Ramos, and H. Shroff, "Inverted selective plane illumination microscopy (iSPIM) enables coupled cell identity lineaging and neurodevelopmental imaging in Caenorhabditis elegans," Proc Natl Acad Sci **108**(43), 17708-17713 (2011).
16. J. Huisken, and D. Y. R. Stainier, "Even fluorescence excitation by multidirectional selective-plane illumination microscopy (mSPIM)," Opt. Lett. **32**(17), 2608–2610 (2007).
17. L. A. Royer, W. C. Lemon, R. K. Chhetri, Y. Wan, M. Coleman, E. W. Myers, and P. J. Keller, "Adaptive light-sheet microscopy for long-term, high-resolution imaging in living organisms," Nat. Biotechnol. **34**(12), 1267–1278 (2016).
18. T. V.Truong, W. Supatto, D. S. Koos, J. M. Choi, and S. E. Fraser. "Deep and fast live imaging with two-photon scanned light-sheet microscopy." Nature Methods **8**(9), 757-760 (2011).
19. C. Dunsby, "Optically sectioned imaging by oblique-plane microscopy," Opt. Express **16**(25), 20306–20316 (2008).
20. S. Kumar, D. Wilding, M. B. Sikkel, A. R. Lyon, K. T. MacLeod, and C. Dunsby, "High-speed 2D and 3D fluorescence microscopy of cardiac myocytes," Opt. Express **19**(15), 13839-13847 (2011).
21. M. B. Sikkel, S. Kumar, V. Maioli, C. Rowlands, F. Gordon, S. E. Harding, A. R. Lyon, K. T. MacLeod, and C. Dunsby, "High speed sCMOS-based oblique plane microscopy applied to the study of calcium dynamics in cardiac myocytes," J. Biophotonics **9**(3), 311-323 (2016).
22. M. B. Bouchard, V. Voleti, C. S. Mendes, C. Lacefield, W. B. Grueber, R. S. Mann, R. M. Bruno, and E. M. Hillman, "Swept confocally aligned planar excitation (SCAPE) microscopy for high-speed volumetric imaging of behaving organisms," Nature Photon. **9**(2), 113–119 (2015).
23. Y. Shin, D. Kim, and H. S. Kwon, "Oblique scanning two-photon light sheet fluorescence microscopy for rapid volumetric imaging," J. Biophotonics e201700270 (2018).
24. T. Li, S. Ota, J. Kim, Z. J. Wong, Y. Wang, X. Yin, and X. Zhang, "Axial plane optical microscopy," Sci. Rep. **4**, 7253 (2014).
25. J. W. Goodman, *Introduction to Fourier optics* (Roberts and Company Publishers, 2005).
26. E. J. Botcherby, R. Juskaitis, M. J. Booth, and T. Wilson, "Aberration-free optical refocusing in high numerical aperture microscopy," Opt. Lett. **32**(14), 2007-2009 (2007).
27. E. Arthur, N. Amodaj, K. Hoover, R. Vale, and N. Stuurman, "Computer control of microscopes using µManager," Current protocols in molecular biology 14-20, (2010).
28. A. D. Edelstein, M. A. Tsuchida, N. Amodaj, H. Pinkard, R. D. Vale, and N. Stuurman, "Advanced methods of microscope control using µManager software," Journal of biological methods, **1**(2), (2014).
29. E. H. W. Meijering, W. J. Niessen, and M. A. Viergever, "Quantitative Evaluation of Convolution-Based Methods for Medical Image Interpolation," Med. Image Anal. **5**(2), 111-126 (2001).
30. C. A. Schneider, W. S. Rasband, and K. W. Eliceiri, "NIH Image to ImageJ: 25 years of image analysis," Nature Methods **9**(7), 671-675, (2012).
31. X. Xiao, V. F. Geyer, H. Bowne-Anderson, J. Howard, and I. F. Sbalzarini, "Automatic optimal filament segmentation with sub-pixel accuracy using generalized linear models and B-spline level-sets," Med. Image Anal. **32** 157–172, (2016).
32. J. Schindelin, I. Arganda-Carreras, E. Frise, V. Kaynig, M. Longair, T. Pietzsch, S. Preibisch, C. Rueden, S. Saalfeld, B. Schmid, and J. Y. Tinevez, "Fiji: an open-source platform for biological-image analysis," Nature Methods **9**(7), 676-682 (2012).
33. L. A. Royer, M. Weigert, U. Günther, N. Maghelli, F. Jug, I. F. Sbalzarini, and E. W. Myers, "ClearVolume: open-source live 3D visualization for light-sheet microscopy," Nature Methods **12**(6), 480-481 (2015).
34. J. Shin, H. C. Park, J. M. Topczewska, D. J. Mawdsley, and B. Appel, "Neural cell fate analysis in zebrafish using olig2 BAC transgenics," Methods Cell Sci. **25**(1-2), 7-14 (2003).
35. C. Satou, Y. Kimura, H. Hirata, M. L. Suster, K. Kawakami, S. I. Higashijima, "Transgenic tools to characterize neuronal properties of discrete populations of zebrafish neurons," Development **140**(18), 3927-3931 (2013).
36. T. R. Thiele, J. C. Donovan, and H. Baier, "Descending control of swim posture by a midbrain nucleus in zebrafish," Neuron **83**(3), 679-691 (2014).
37. E. Baumgart, and U. Kubitscheck, "Scanned light sheet microscopy with confocal slit detection," Opt. Express **20**(19), 21805-21814 (2012).
38. J. G. McNally, T. Karpova, J. Cooper, and J. A. Conchello, "Three-dimensional imaging by deconvolution microscopy," Methods **19**(3), 373-385 (1999).
39. F. O. Fahrbach, V. Gurchenkov, K. Alessandri, P. Nassoy, and A. Rohrbach, "Light-sheet microscopy in thick media using scanned Bessel beams and two-photon fluorescence excitation," Opt. Express **21**(11), 13824-13839 (2013).
40. P. J. Dwyer, C. A. DiMarzio, and M. Rajadhyaksha, "Confocal theta line-scanning microscope for imaging human tissues," Appl. Opt. **46**(10), 1843-1851 (2007).

41. H. Yu, V. Voleti, K. Patel, W. Li, M. A. Shaik, and E. M. Hillman, "Two-photon Swept Confocally Aligned Planar Excitation Microscopy (2P-SCAPE)," In Novel Techniques in Microscopy, (Optical Society of America, 2017) pp NW4C-3.
42. B. C. Chen, W. R. Legant, K. Wang, L. Shao, D. E. Milkie, M. W. Davidson, C. Janetopoulos, X. S. Wu, J. A. Hammer, Z. Liu, B. P. English, Y. Mimori-Kiyosue, D. P. Romero, A. T. Ritter, J. Lippincott-Schwartz, L. Fritz-Laylin, R. D. Mullins, D. M. Mitchell, J. N. Bembenek, A.-C. Reymann, R. Böhme, S. W. Grill, J. T. Wang, G. Seydoux, U. S. Tulu, D. P. Kiehart, and E. Betzig, "Lattice light-sheet microscopy: imaging molecules to embryos at high spatiotemporal resolution," Science **346**(6208), 1257998 (2014).
43. T. Vettenburg, H. I. Dalgarno, J. Nylk, C. Coll-Lladó, D. E. Ferrier, T. Čižmár, F. J. Gunn-Moore, and K. Dholakia, "Light-sheet microscopy using an Airy beam," Nature Methods. **11**(5), 541 (2014).

## 1. Introduction

The field of biological imaging is driven by the pressing need for new techniques that offer higher resolution, faster acquisition speed, and deeper imaging capabilities. Modern neuroscience experiments frequently require *in vivo* or even whole organism imaging, at optimized speed, depth, and resolution. Steric access is also essential, so that additional modalities, from electrophysiology to sophisticated virtual reality systems can be integrated along with imaging [1]. This challenge requires overcoming scattering, absorption, and photobleaching, associated with many biological samples. While super-resolution microscopy approaches STED [2], STORM [3], PALM [4], and SIM [5] provide the greatest resolution, they compromise on imaging speed, limiting many functional imaging applications. Two-photon imaging has become the gold-standard for deep tissue high resolution imaging [6], yet point scanning approaches suffer from relatively slow imaging speed, especially in volumetric imaging. Light-field microscopy offers the fastest volumetric imaging [7], limited only by camera frame rate, and some implementations of these techniques have been able to measure dynamic fluorescence signals from deep inside scattering tissues [8]. However, image visualization in light-field microscopy is computationally heavy, and live monitoring of samples at subcellular resolution has not yet been attained.

Light-sheet microscopy, forgotten for over a century [9], has recently re-emerged as a powerful imaging technique [10]. It provides improved resolution [11], with reduced effect of scattering [12], and high-speed functional imaging capability [13]. A conventional light-sheet microscope consists of an illumination arm arranged orthogonally to an upright detection arm [9,10]. The illumination arm relies on a cylindrical lens to focus a collimated beam to form a sheet of light, providing optical sectioning of samples. The detection arm forms a magnified image of the optically sectioned plane. The axial resolution of a light-sheet microscope depends on the detection objective numerical aperture (NA) and the light-sheet thickness. Thus, it is possible to attain higher resolution imaging by thinning the imaging light-sheet [11]. While optical sectioning improves axial resolution, it also reduces photobleaching and out-of-focus background signal, circumventing the effects of scattering and improving image contrast. Several designs of light-sheet microscopes have been developed, including OpenSPIM [14], inverted SPIM [15], and multidirectional SPIM [16]. The new advances have also enabled adaptive long-term live sample imaging capabilities [17]. Introduction of a fast scanning mirror has allowed for creation of light-sheet for two-photon excitation of the sample, providing deeper and much higher resolution in scattering samples using light-sheet microscopy [18], while still maintaining reduced sample bleaching. Despite many recent developments, light-sheet microscopy remains limited in size and orientation of the imaged sample, due to the steric hindrance associated with designs using two to four objectives to create the light-sheet and image optically sectioned sample plane. Thus, *in vivo* imaging of larger organisms remains out of reach for conventional multi-objective light-sheet systems.

One approach to overcome the current constraints on light-sheet microscopy relies on single objective based light-sheet microscopy system designs. In 2008, Dunsby introduced a single front facing objective based oblique plane microscopy (OPM), where one high NA objective is used for both illuminating an oblique plane in the sample and imaging it [19]. OPM system uses three microscope sub-systems arranged sequentially, which helps correct the aberrations introduced by oblique illumination, and it employs rotation optics to focus the emission from the illuminated plane on a camera sensor. Later, rapid volumetric imaging in OPM systems was achieved by piezo-assisted scanning of the second objective along axial direction [20,21]. Recently, SCAPE microscopy introduced an alternative way for rapid scanning by introducing a polygonal scan mirror between the first two microscopy sub-systems of OPM [22]. SCAPE simplified the OPM optical arrangement by moving the excitation beam arrangement from the tight space between two objectives to the larger space between the two tube-lenses, allowing for lateral rather than axial scan direction of the light-sheet. While OPM and SCAPE systems support rapid volumetric imaging, both are mostly limited to one-photon light-sheet imaging. Moreover, where OPM system implements a somewhat inconvenient design to launch and scan the light-sheet into sample, SCAPE suffers from scan-position dependent tilt in the light-sheet. Even the alternative proposed SCAPE design with two synchronized planar mirrors for the scanning-descanning arrangement was expected to maximize the detection numerical aperture [22], but it would not resolve the light-sheet orientation variance inherent in the scanning architecture. This limitation creates a scan-position dependent point spread function (psf) and makes the exact 3D reconstruction of scanned volume computationally heavy [22,23]. Very recently, a modified form of axial plane optical microscopy [24], called OS-2P-LSFM [23] has been proposed as an approach to overcome scan position dependent tilt in single front facing objective based oblique plane light-sheet microscopy. This approach utilizes a refractive glass window as a scanning element to help maintain constant light-sheet sweep angle, but it is limited to low-tilt angle (nearly axial) light-sheet architecture. In addition, heavy beam clipping at intermediate objective [24] leads to low axial resolution, and the spherical and chromatic aberrations caused by the glass window become severe constraints beyond small scan ranges. Even the advantage of implementing two-photon light-sheet is partly compromised due to limited axial resolution of the setup.

In this work, we address the existing constraints with single front facing objective based light-sheet architecture. First, we use a single plane mirror based scanning architecture which is not only simpler than polygon scan mirror based SCAPE implementation [22] but also solves the issue of scan-position dependent tilt in the generated light-sheet. Next, we bring two-photon light-sheet imaging capability to a single-front facing objective based system. Our scanned oblique plane illumination (SOPi) microscopy system's streamlined design allows us to seamlessly integrate both one-photon and two-photon light-sheet imaging capability in the same system, enabling easy switching between excitation modes while imaging a region of interest in a sample. We use our system for volumetric imaging in mouse brain sections and for *in vivo* structural and functional imaging of behaving zebrafish larvae.

## 2. Experimental setup and volume reconstruction

Here we describe the optimized scanning architecture behind our SOPi system, its complete optical construction, and steps involved in reconstructing the volume data acquired with it.

*2.1 Scanning arrangement*

At the heart of SOPi system lies its simple yet optimized optical scanning geometry. Figure 1 illustrates the main idea behind the creation of SOPi system through the introduction of a scanning architecture to OPM design. In OPM, a cylindrical lens focuses a laser beam to form a focus line along the y-axis at an offset position to the back focal plane (BFP) of MO1, in order to produce oblique light-sheet illumination in the sample volume. This light-sheet is tilted in the y-z plane but remains parallel to the x-axis. The illuminated oblique sample plane is then re-imaged at its conjugate oblique plane in front of MO2. This intermediate image plane is then magnified by MO3-L5 microscope to be imaged on a camera. The shared on-axis image location between L1 and L4 lenses allows us to insert a scanner in this plane, which would ideally shift rays without introducing any additional tilt. We start by looking for the simplest scanning geometry consisting of two identical lenses (L2, L3) and a galvo scanner based plane mirror (G1), arranged as shown in the inset of Fig. 1(a). Since G1 lies in the Fourier plane of both entry and exit ports, these ports become conjugate image planes to each other by dual optical Fourier transform operation [25]. So, this scanning arrangement can be inserted at the plane marked as 'scanning plane' of OPM setup without affecting its normal operation. Figure 1(b) shows this modified setup, obtained by introduction of scanning geometry, which we refer to as SOPi setup for its intended use for obtaining scanned oblique plane illumination of excitation beam. SOPi uses G1 to scan the light-sheet along the y-axis without causing any change in its tilt angle (in the y-z plane) and de-scans the generated fluorescence signal to yield a stationary intermediate image of the illuminated plane.

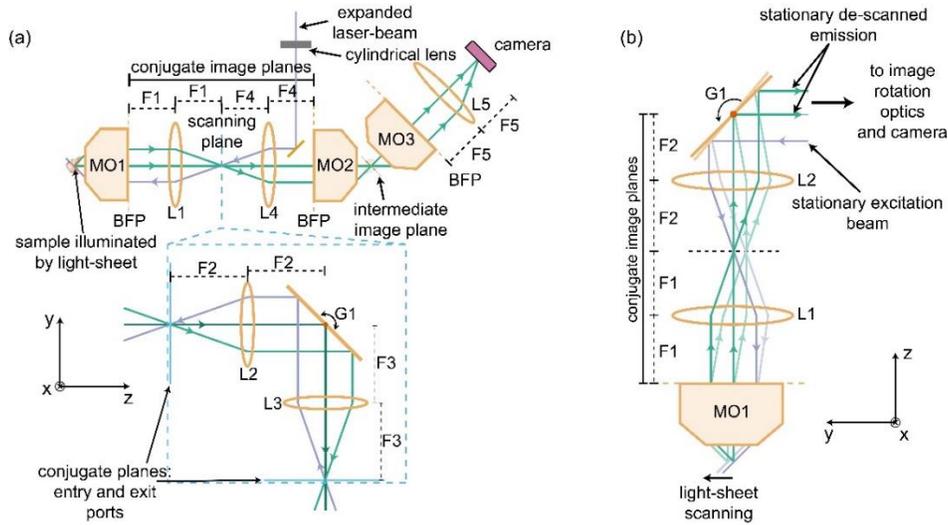

Fig. 1. Scanning architecture of SOPi. (a) SOPi is assembled by integrating a planar scan mirror based scanning geometry (shown in inset) into the OPM design. MO: microscope objective, BFP: back focal plane. (b) Part of assembled SOPi design showing the intended action of the scan mirror to control the scanning of light-sheet without a change in tilt while simultaneously de-scanning the fluorescence signal to provide a stationary emission beam.

From the principle of Fourier optics, we know that optical equivalence of Shift theorem assures that a tilt in the back focal plane of a lens becomes a perfect shift without any change in tilt in the front focal plane [25]. It follows that an excitation beam incident at on-axis location in the setup [Fig. 2(a)], with its point of reflection (pivot point) exactly at the back focal plane of L2, undergoes a perfect shift during scanning. Yet this arrangement would provide an axial, rather than an oblique, light-sheet. The generation of an oblique light-sheet relies on off-axis incidence of excitation beam at BFP of MO1. This requirement can be met by either shifting G1 or the excitation beam itself as shown in Fig. 2(b) and 2(c) respectively. While shifting G1 perturbs the conjugate plane relationship between entry and exit ports of the scanning arrangement [inset Fig. 1(a)], shifting the excitation beam causes the pivot point to move away from the back focal plane of L2. Either condition would deteriorate scanning/de-scanning performance. Given that the amount of offset required (a few mm) is a fraction of the focal length of L2 (here, 100 mm), it follows that shifting G1 to incur the desired offset [Fig. 2(b)] is a superior choice, since it maintains the pivot point of the excitation beam at the back focal plane of L2. Then subsequent optics can be slightly adjusted to compensate for the change in location of exit port in the scanning arrangement. To validate this reasoning, we performed ray tracing based optical modeling (using OpticStudio, Zemax LLC) of all three arrangements shown in Fig 2.

It is sufficient, in theory, to simulate the scanning of a thin excitation beam which lies in the plane of the diagram. Once the behavior of this beam is established, it can be generalized to other beams within the light-sheet. This is possible because scan-mirror G1 is based on a single axis galvo, and its rotation along the x-axis affects the component of rays lying exclusively in the plane of rotation (y-z plane). We set up our simulation using L2 (Achromat doublet lens, f = 100 mm, AC508-100-A-ML, Thorlabs), a scan mirror, a numerical plane detector, and 3.54 mm for the numerical value of the shift in G1 position/beam-offset as per Fig. 2(b) and 2(c). The calculation of 3.54 mm as the required offset is shown in the following section. In our simulation, we recorded the beam-position and beam-tilt angle ($\varphi$) at the detector plane for G1 scan angle ($\theta$) in the range 43°-47° and plotted them in Fig. 2(d) and 2(e). While the position of the beam does not change among the three scan geometries (owing to small scan angle and small offset relative to the focal length of the lens), the tilt angle is highly sensitive to the choice. Surprisingly, the arrangement based on an offset of the beam as shown in Fig. 2(c) shows a constant tilt during the beam scanning process, and hence it is the optimal way to introduce the required offset. This result dramatically simplifies the experimental setup of SOPi, as it can be arranged by aligning all the optical elements (including galvo scanner) between MO1 and MO2 along a common optical-axis and then introducing the desired offset in the incoming excitation beam. This result also describes why a scanning arrangement employing polygon mirror, which can only favor scanning geometry of Fig. 2(b), is a suboptimal choice when aiming for constant tilt scanning.

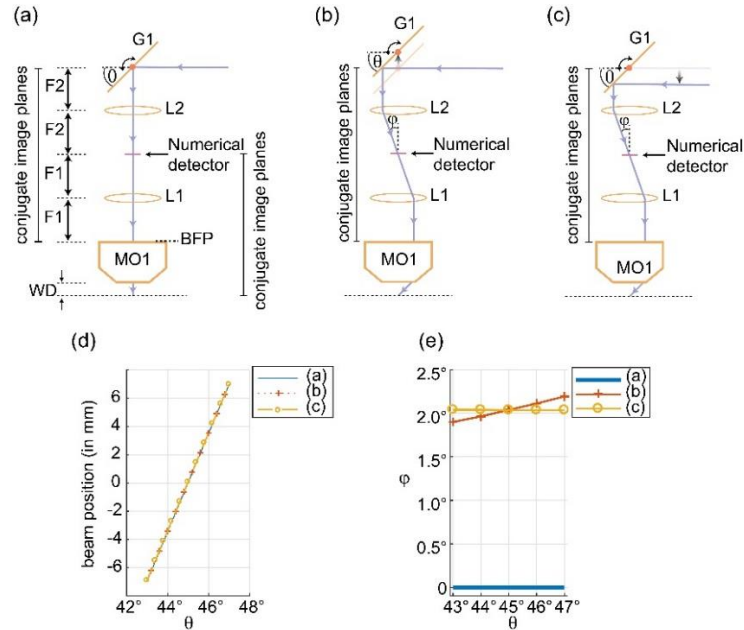

Fig. 2. Optimization of scanning geometry. (a) Scanning architecture with on-axis excitation beam. (b) Shifting the scan-mirror provides an offset in the incident beam to produce an oblique light-sheet. The point of reflection is centered at back-focal plane of L2 and shifts around it as G1 scans. (c) Shifting of excitation beam from its zero-position (as shown in (a)) to produce desired oblique light-sheet. The point of reflection is no longer centered at back-focal plane of L2 and shifts as G1 scans. (f) Ray-tracing based numerically measured scanned beam-position at the numerical detector plane. (g) Numerically measured scanned beam-tilt at the detection plane.

*2.2 SOPi optical layout*

Our integrated SOPi microscope [Fig. 3(a)], makes use of two excitation sources. For 1P excitation, we used a low-cost laser diode ($\lambda$ = 462 nm, L462P1400MM, Thorlabs) driven by a commercial benchtop variable power supply (Tekpower TP3005T). For two-photon excitation, we used a tunable ultrafast laser (680-1300 nm, InSight DeepSee, Spectra-Physics). We collimated the laser diode emission by passing it through an achromatic doublet lens (L8, f = 50 mm, AC254-050-A-ML, Thorlabs). This collimated beam was then passed through a slit-aperture (VA100, Thorlabs) and a plano-convex cylindrical lens (L9, f = 50 mm, LJ1695RM-A, Thorlabs) to focus the beam to a line. Reflection from a dichroic mirror (DM2, 470 nm single-edge long-pass, FF470-Di01-25x36, Semrock) allowed the beam to be focused on a galvanometer mounted planar silver mirror (G1, QS-12, 10 mm clear aperture, Nutfield Technology), connected to a driver board (QD4000, Nutfield Technology). The line focus orientation was perpendicular to the galvanometer's axis of rotation, and the origin position of scan mirror was set at 45° to the incoming beam. Along the reflected path from the scan mirror, we used an imaging setup consisting of two achromatic doublet lenses (L1, f = 200 mm, AC508-200-A-ML & L2, f = 100 mm, AC508-100-A-ML, Thorlabs) to re-image this focused line onto the back-focal plane of the main microscope objective (MO1, 20x, 1.0 W, XLUMPLFLN20XW, Olympus). The objective performs a Fourier transformation of the beam to produce a light-sheet in the sample volume. As expected from the Fourier transform properties, the light-sheet orientation is perpendicular to the line-shaped focus at the back-focal plane of the objective. Then, rotation of the scan mirror gives rise to pure translation of the light-sheet in front of the objective with no change in the tilt angle. The same main objective collects the fluorescence signal from the sample, which follows the path of the excitation beam backwards to get reflected off the scan mirror. We use a dichroic mirror (DM1, 640 nm single-edge, long-pass, FF640-FDi01-25x36, Semrock) to reflect the emitted fluorescent signal to another two-lens relay system consisting of achromatic doublet lenses (L3, f = 100 mm, AC254-100-A-ML & L4, f = 150 mm, AC254-150-A-ML, Thorlabs) which images the center of the scan mirror onto the back-focal plane of a dry super-achromat microscope objective (MO2, 20x, NA 0.75, UPLSAPO20X, Olympus). This arrangement sets the working distance of the two objectives (MO1 and MO2) as conjugate image planes of one other. Moreover, the choice of MO1-L1, L2-L3 and MO2-L4 is made in such a way that intermediate image plane has the same lateral and axial magnification [18,26].

The same galvanometer-mounted mirror (G1) responsible for scanning the light-sheet also de-scans the fluorescence signal to provide a stationary intermediate image. We introduced an offset (3.54 mm as calculated below) in the incoming excitation beam by shifting the laser-diode, collimating lens (L8), and slit-aperture arrangement so that the line-shaped focused beam at back-focal plane of the objective falls at an off-axis position to give rise to a 45° tilted light-sheet in the sample volume. This arrangement produces an intermediate 45° tilted image plane of the oblique light-sheet illuminated sample in front of the dry objective (MO2). We then used a third microscope objective (MO3, 20x, NA 0.45, LUCPLFLN20X, Olympus) along with an achromatic doublet lens (L5, f = 100 mm, AC254-100-A-ML, Thorlabs) and sCMOS camera (Prime 95B, Photometrics) to form a magnified image of this intermediate oblique image plane on the camera. The third objective is oriented at 45° with respect to the principal axis of the second objective [Fig. 3(a)], so that the intermediate image plane is located exactly at the working distance of the third microscope objective. For this precise positioning, we used the fine adjustment manual translation stage (SM1Z, Thorlabs) to move the third microscope objective in the position to enable accurate imaging of the oblique intermediate image plane on the stationary camera. In this setup, rotation of the scan mirror enables scanning in sample volume, and the scanned oblique plane always remains in focus on the camera.

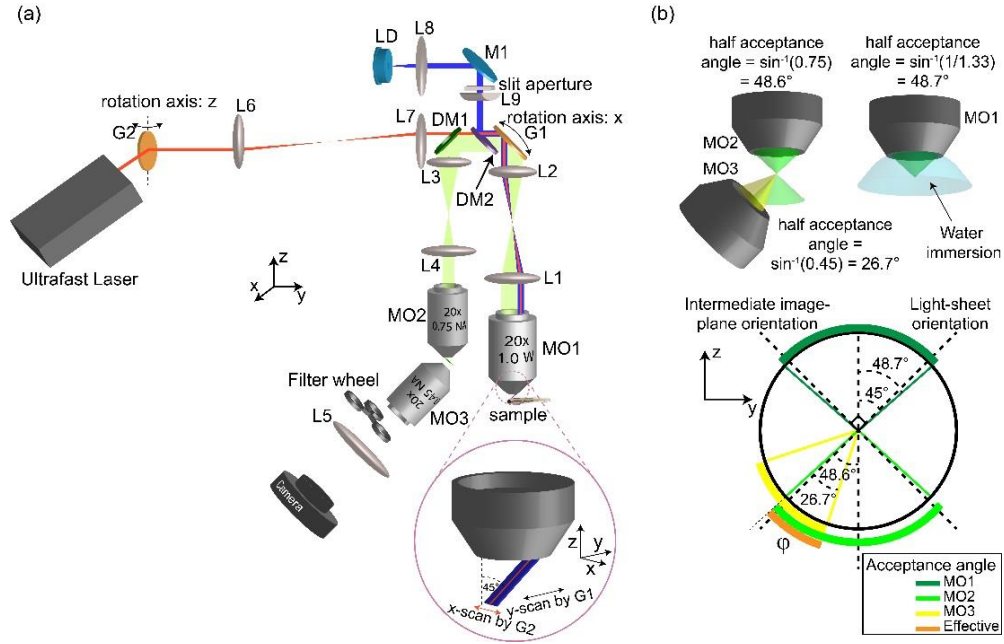

Fig. 3. SOPi system. (a) Schematic diagram describing the full optical layout of SOPi. Inset shows the scanning arrangement to create and sweep light-sheet in sample volume. MO: microscope objective, DM: dichroic mirror, LD: laser diode, M: mirror. (b) Calculation of effective acceptance angle of SOPi system.

We then added 2P light-sheet imaging capability analogous to DSLM [12]. To maintain a small illumination numerical aperture and therefore large Rayleigh range, we did not expand the laser beam. We reflected the ultrafast laser beam with a galvanometer-mounted plane mirror (G2, GVSM001, Thorlabs) to produce a light-sheet by fast scanning. We used a relay setup consisting of two identical achromatic lenses (L6, L7 f = 100 mm, AC254-100-B-ML, Thorlabs) to image scan mirror G2 onto another scan mirror G1, responsible for scanning the light-sheet, as described above. We oriented the two-photon beam scanning galvanometer G2 so that its rotation axis is orthogonal to another galvanometer's (G1's) rotation axis [Fig. 3(a)]. Then, we introduced the same offset (3.54 mm as calculated below) to the laser beam forcing the 2P light-sheet to undergo a 45° tilt in front of the main objective, co-aligning with the 1P light-sheet orientation. During imaging experiments, we controlled both galvo scanners by a custom MATLAB GUI. This GUI generated ramp voltage signal output, with the help of a data acquisition card (DAQ, PCIe-6321, National Instruments, 2 analog output channels). Users control the amplitude and frequency of the ramp signals, which directly translates into mirror scan range and scanning speed, respectively. We evaluated the relationship between applied voltage from the DAQ, tilt-angle of scan mirrors, and the actual physical sweep distance moved by the light-sheet in the sample volume for both galvanometers in the setup. We used the corresponding scaling factors to enable the selection of scanning distance (in µm) and scanning time/frequency (in seconds/Hz) directly from the MATLAB GUI. For camera control and image acquisition we used µManager [27,28], an open source microscopy control software. The scanning speed of light-sheet and camera's acquisition frame-rate determine voxel depth. Numerically, voxel depth is obtained from a single sweep of the scanned volume by taking a ratio of scan-range (in µm) and total number of frames. For example, scanning a 500 µm range of a sample in 10 seconds at 50 fps camera speed yields voxel depth = scan-range ÷ number of camera frames = 500 µm ÷ 500 = 1 µm.

Next, we describe the calculation of required offset to generate the desired oblique light-sheet. The oblique light-sheet generation is based on the Fourier transforming property of the optical lens, and the numerical aperture of the lens (water immersion objective MO1) limits the attainable tilt to a maximum value of $\sin^{-1}(1/1.33) = 48.75°$. To avoid clipping the beam at the edge of the aperture, we set the target tilt angle to a slightly smaller value of 45°. The required offset to get this desired 45° oblique light-sheet is readily calculated. From Fig 2(c), we note that L1-MO1 combination forms a microscope with lateral or angular magnification of $20 \times 200/180 = 200/9$. Thus, a beam travelling from numerical detector plane to the sample plane in front of MO1 is demagnified in its spatial position, and magnified in its angular tilt by the same factor of 200/9. For a 45° tilt in sample plane, this requires the beam tilt at numerical detector plane to be $45° \times 9/200 = 2.025°$. Given the focal length of L2 (f = 100 mm), the required offset of incident beam is calculated as beam-offset = focal length × tan(angular tilt) = 100 mm × tan(2.025°) ≈ 3.54 mm.

One of the main drawbacks of placing three objectives sequentially is the limit on the effective numerical aperture of the overall system. The effective numerical aperture is obtained from maximum cone angle the system can effectively gather light from and deliver it to an image-forming element. Figure 3(b) shows how the third objective in the SOPi setup becomes the main limiting factor in defining the overall system numerical aperture. The total effective acceptance angle of SOPi in current configuration is $\varphi = 48.6° - 90°/2 + 26.7° = 30.2°$. Hence, effective numerical aperture = $n_{water} \times \sin(\varphi/2) = 1.33 \times \sin(30.2°/2) \approx 0.34$. We can also calculate the effective magnification of SOPi as the product of three individual magnifications of constituent microscope sub-systems (MO1-L1, L4-MO2 and MO3-L5). $M1 = 20 \times 200/180$, $M2 = 1 \div (20 \times 150/180)$ and $M3 = 20 \times 100/180$. Hence, $M = M1 \times M2 \times M3 = 200/150 \times 20 \times 100/180 = 400/27 \approx 14.81$. With camera pixel size of 11 µm × 11 µm it is easy to determine the effective voxel width and height in image space. Voxel width = voxel height = 11 µm ÷ 14.81 ≈ 0.74 µm.

*2.3 Affine transformation for volume reconstruction*

SOPi comes with a unique scanning geometry where a 45° oblique illumination plane is scanned along the perpendicular direction to the optical axis of the microscope objective. As described in Fig. 4, the volume acquired in this geometry cannot be reconstructed by simple stacking of the acquired images.

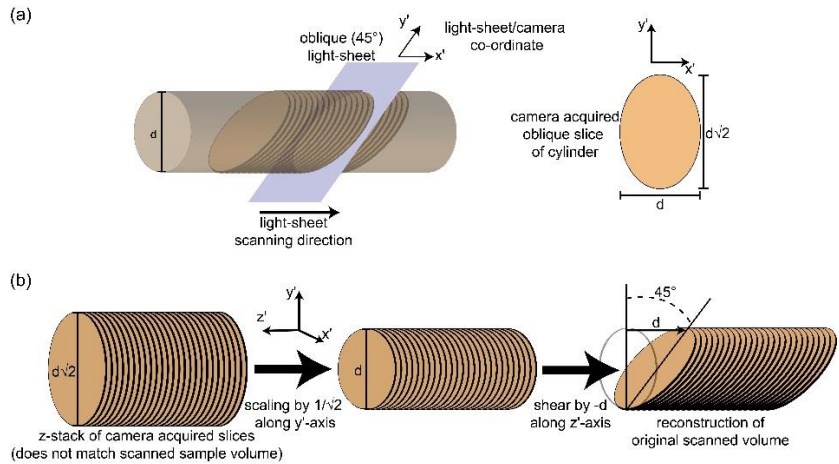

Fig. 4. Affine transformation for correct volume reconstruction. (a) Light-sheet orientation in a cylindrical object and corresponding image section acquired on camera. (b) Geometrical transformations to reconstruct the scanned volume.

The reconstruction depends on combining two geometrical transformations (scaling and shearing) together. A geometrical transformation operation (in 3D Cartesian coordinate) is described by following matrix operation:

$$\begin{bmatrix} x \\ y \\ z \\ 1 \end{bmatrix} = \begin{bmatrix} a_{xx} & a_{xy} & a_{xz} & a_{xt} \\ a_{yx} & a_{yy} & a_{yz} & a_{yt} \\ a_{zx} & a_{zy} & a_{zz} & a_{zt} \\ 0 & 0 & 0 & 1 \end{bmatrix} \cdot \begin{bmatrix} x_i \\ y_i \\ z_i \\ 1 \end{bmatrix}. \quad (1)$$

The combined affine transformation matrix is given by the product of scaling ($M_{sc}$) and shearing ($M_{sh}$) matrices (keeping their order of operation in mind):

$$M_{sh} \times M_{sc} = \begin{bmatrix} 1 & 0 & 0 & 0 \\ 0 & 1 & 0 & 0 \\ 0 & -1 & 1 & 0 \\ 0 & 0 & 0 & 1 \end{bmatrix} \times \begin{bmatrix} 1 & 0 & 0 & 0 \\ 0 & 1/\sqrt{2} & 0 & 0 \\ 0 & 0 & 1 & 0 \\ 0 & 0 & 0 & 1 \end{bmatrix} = \begin{bmatrix} 1 & 0 & 0 & 0 \\ 0 & 1/\sqrt{2} & 0 & 0 \\ 0 & -1/\sqrt{2} & 1 & 0 \\ 0 & 0 & 0 & 1 \end{bmatrix}. \quad (2)$$

We used this affine transformation matrix in transformJ [29], an ImageJ [30] plugin, to perform 3D geometrical transformation of the acquired image data.

## 3. Results and methods

In this section, we describe several experiments performed using both 1P and 2P light-sheets of the SOPi system.

### 3.1 Imaging microbeads

Microbeads embedded into agar gel were used for evaluating the imaging performance of SOPi. We first prepared a 0.5 weight % solution of agarose (LE Analytical Grade, Promega™ V3125) in Milli-Q water. Then, 1 µL solution of 0.5 µm fluorescent microspheres (TetraSpeck™, Thermo Fisher-T7284) was added to 20 mL of agarose solution. The mix was vortexed, heated, and then cooled in a petri dish to form a volume sample. Figure 5(a) and 5(b) show the lateral view of the microbeads imaged using 1P and 2P light-sheet respectively. The lateral view is obtained by performing a maximum intensity projection of geometrically corrected image stacks. The anisotropy apparent in the oval shape of the point spread function (psf) arises due to the elliptical NA of the SOPi system. Insets display enhanced images to illustrate the extended tail of 1P psf (point spread function) arising due to some residual system aberration and the thicker light-sheet. The normalized intensity line-plot through one of the beads [Fig. 5(c)] also demonstrates the superior resolution capability of 2P SOPi. To determine the resolution capability of SOPi, we measured lateral FWHM (full width at half maxima) of seven microbeads using MosaicSuite plugin [31] in Fiji [32] and found 1P FWHM and 2P FWHM to be 1.30 µm (standard deviation 0.09 µm) and 1.16 µm (standard deviation 0.06 µm), respectively.

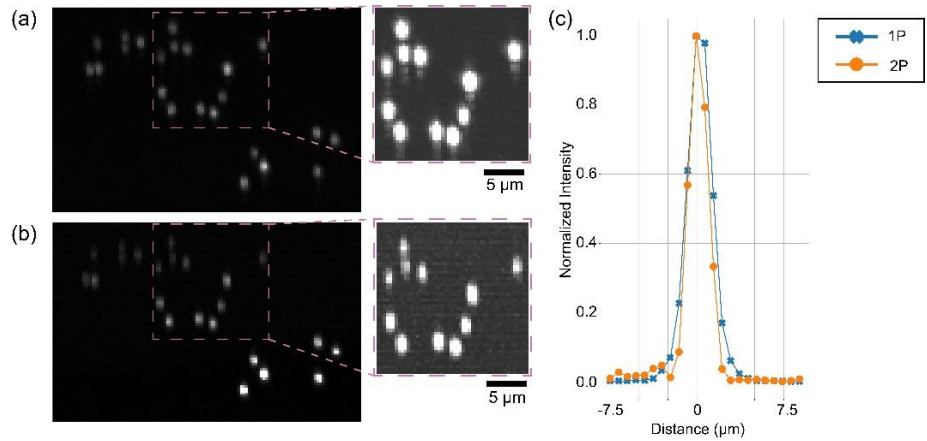

Fig. 5. Imaging microbeads for estimating resolution. (a) 1P SOPi microscopy of microbeads. (b) 2P SOPi microscopy of microbeads. Insets in (a) and (b) show enhanced images to help see low intensity artifacts. (c) Normalized intensity line plot through one microbead.

## 3.2 Imaging mouse brain slice

Using SOPi we imaged a fixed, not optically cleared 1 mm thick section of Thy1-GFP transgenic mouse hippocampus [Fig. 6(a)]. We used laser diode assisted 1P SOPi system and acquired a sequence of 600 images in 6 seconds at 10 ms exposure time through 850×325×500 µm$^3$ volume within the sample. A 3D reconstruction of the volume was then obtained [Fig. 6(b)] by z-stacking all acquired frames and visualizing with ClearVolume plugin [33] in Fiji.

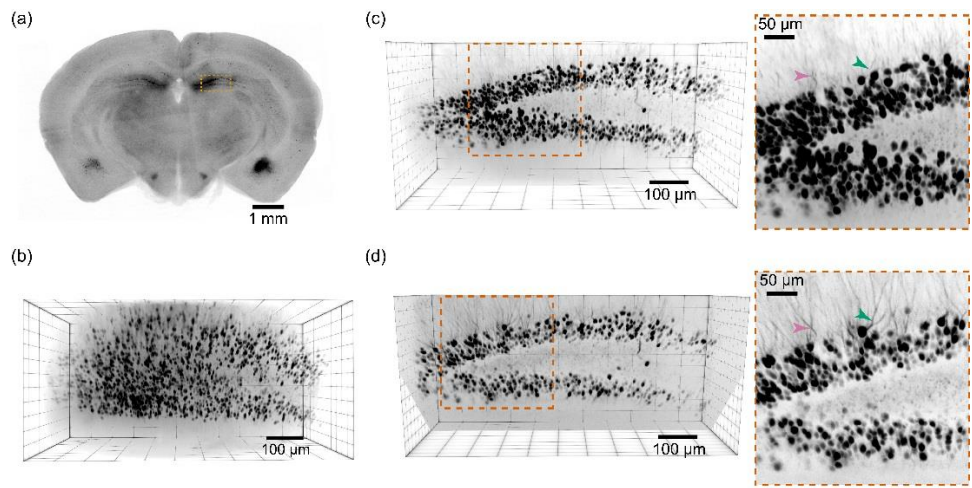

Fig. 6. Imaging mouse brain slice using SOPi. (a) A widefield fluorescence image of 1 mm thick slice of thy1GFP adult mice along with a highlighted area of the hippocampus imaged under SOPi setup. (b) Volumetric reconstruction of SOPi acquired 1P light-sheet images by z-stacking frames. (c) Affine transformed 3D reconstruction of 1P light-sheet scanned volume which matches the view of the dentate gyrus as expected from (a). The inset shows a zoomed in version to illustrate finer dendritic details of dentate gyrus granule neurons. (d) Affine transformed 3D reconstruction of the same volume scanned by 2P light-sheet on SOPi setup. Inset image demonstrates superior dendritic imaging compared to their 1P light-sheet imaged copy in (c). Two arrows (pink, green) facilitate direct comparison of the same dendritic region in both 1P and 2P imaging.

As expected, this produces a geometrically distorted 3D reconstruction of original volume. To correct this distortion, we applied affine transformation to the stacked data using transformJ plugin in imageJ/Fiji. The resulting volume is displayed in Fig. 6(c) and visualization 1 shows true 3D perspective of the scanned volume. Neurons and their dendrites are easily tracked throughout the scanned volume in the slice. To perform 2P light-sheet imaging, we switched to an ultrafast laser tuned to 910 nm. We adjusted the laser power to obtain well exposed images at 50 ms exposure time while minimizing bleaching, and adjusted light-sheet scanning mirror range and speed to obtain 600 frames to scan through 750×270×500 µm$^3$ volume in 30 seconds. Repeating the same process of stacking frames followed by affine transformation, we obtained a volume reconstruction as shown in Fig. 6(d) and visualization 2. 2P light-sheet imaging has superior structural imaging capability that comes at the cost of speed, as the 2P fluorescence cross section is much smaller than 1P fluorescence cross section for illumination by a low numerical aperture excitation beam.

### *3.3 Imaging zebrafish*

SOPi is well-adapted for imaging live and behaving zebrafish, so we next performed both functional and structural *in vivo* imaging of zebrafish larvae. For structural imaging, we targeted the densely labelled cerebellum region of a GFP expressing fish brain (5-day-post-fertilization nacre Tg(Olig2:GFP) [34] larvae).

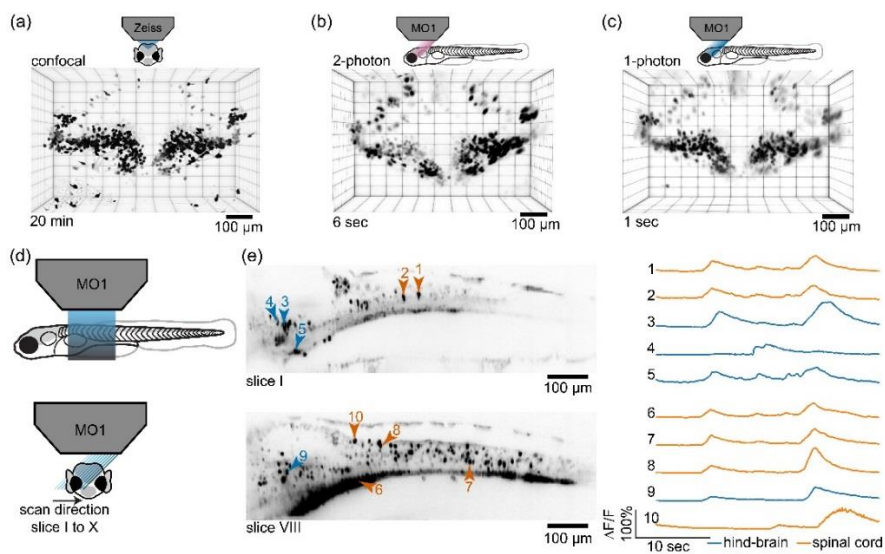

Fig. 7. Imaging zebrafish larvae. (a) A high resolution confocal imaging of zebrafish cerebellum in nacre Tg(Olig2:GFP) fish acquired in 20 min. (b) The same cerebellar region imaged with 2P SOPi setup in 6 seconds. (c) The same cerebellar region imaged using 1P SOPi setup in 1 second. (d) Schematic diagram showing the arrangement for rapid volumetric GCaMP imaging of Tg(VGlut2a:Gal4;UAS:GCaMP6s) zebrafish hind-brain and spinal cord using a fast scanning 1P light-sheet in the SOPi setup. The volume scan consists of 10 segments, covered at rate of 100 fps leading to 10 VPS scan speed. (e) Left, GCaMP fluorescence in a subset of active cells during spontaneous activity, shown as standard deviation based intensity projections of the frames corresponding to slice position I and VIII in scanned volume (I-X). Right, GCaMP imaging traces corresponding to neurons 1-10 in optical sections I and VIII (30 sec).

A scanned region of 450×300×200 µm$^3$ was sufficient to contain the entire cerebellum. Even with slow high resolution scanning, it took only 6 seconds and 1 second to image this volume with 2P and 1P light-sheets respectively. Using 1P light-sheet, with modest compromise on resolution along the scan direction, we could image same volume in 1/10th of a second (data not shown). We stacked the captured frames followed by affine transformation to get the 3D volume reconstruction of the scanned volume, as shown in Fig. 7(b), 7(c) and visualization 4, 5. To compare the imaging quality of SOPi with conventional imaging modalities we scanned the same volume in the same fish on a Zeiss LSM 710 confocal microscope (~20 minutes) and reconstructed the 3D volume as shown in Fig. 7(a) and visualization 3. Comparing the reconstructions illustrates that SOPi can image most of the cell bodies, even when they are densely packed together, in a small fraction of time compared to point-scanning confocal imaging.

Next, we performed rapid volumetric calcium imaging on a 5-day-post-fertilization GCaMP6s-expressing zebrafish larvae (Tg(VGlut2a:Gal4;UAS:GCaMP6s) [35,36]. For this we used 1P light-sheet from SOPi and imaged a volume section covering the hind brain and spinal cord of fish spanning 850×300×50 µm$^3$. In this volume section, we imaged GCaMP6s-expressing neurons during spontaneous activity in immobilized larvae for 30 seconds at 10 volumes per second rate with 100 frames per second capture rate at 9 ms exposure time on camera. The scan direction of 50 µm was optically sub-divided into 10 segments, such that each segment was imaged at a constant 10 frames per second for the duration of recording. Visualization 6 shows the imaging from each of these 10 segments. Out of few hundred active neurons observed in the scanned volume, we plotted calcium influx response over time as ΔF/F for 10 selected cells [Fig. 7(e)]. The 4D (3D volume + time) rendering of this rapid 10 volumes per second recording is presented in visualization 7.

*3.4 Animal procedures*

Animals were handled according to protocols approved by the Northwestern University Animal Care and Use Committee. Young adult male Thy1-GFP mice (postnatal day ~40, stock # 007788, Jackson laboratory, Bell Harbor, ME) were used in this study. Mice were housed under a 12h light–dark cycle, with food and water available *ad libitum*. For preparing brain slices, mice were deeply anaesthetized with isoflurane and transcardially perfused with 4% paraformaldehyde (PFA) in 0.1 M phosphate buffered saline (PBS). Brains were post-fixed for 2-5 days at -4°C, prior to sectioning. For thick brain slice imaging, tissue containing the hippocampus was sectioned coronally at 1000 µm on a Vibratome (Leica Instruments, Nussloch, Germany), mounted onto Superfrost Plus slides (ThermoFisher Scientific, Waltham, MA), and coverslipped under glycerol:Tris buffered saline (3:1).

Fish were raised and maintained at 28.5°C in an in-house breeding facility. For imaging zebrafish (Danio rerio), experiments were performed in 5-7-day-post-fertilization (dpf) zebrafish larvae. At these age, fish are freely swimming. For structural imaging of neurons, 5 dpf nacre Tg(Olig2:GFP) [34] larvae were used. For calcium imaging experiments, 5-7 dpf Tg(VGlut2a:Gal4;UAS:GCamP6s) [35,36] zebrafish larvae were used. These were bathed in 0.003% 1-Phenyl-2-thiourea (PTU) starting at 18 hours post fertilization, to prevent the formation of melanophores. For all experiments, larvae were first anesthetized in a 0.02% solution of tricaine methanesulfonate (MS-222) and then immersed in 1 mg/ml α-bungarotoxin for 2-3 minutes to prevent muscle activation and movement artifacts. The larvae were then embedded in 1.4% low melting point agarose (Invitrogen) in a glass bottomed Petri dish and then covered in anesthetic-free 10% Hank's solution.

## 4. Discussion and conclusion

SOPi with its integrated 1P and 2P imaging capability is a valuable and potentially broadly applicable single front facing objective based light-sheet system. While 1P SOPi allows for rapid volumetric imaging exceeding 10 volumes per second, 2P imaging has better imaging capability for imaging light-scattering samples. Longer excitation wavelength of 2P light-sheet provides better resolution with no shadow artifacts, when compared to its 1P counterpart. With 2P light-sheet, the improved resolution comes at the cost of reduced speed. Nevertheless, as a line-scan approach it is still an order of magnitude faster than point scanning approaches like confocal microscopy or conventional 2-photon laser scanning microscopy. For a given sample, 2P volume scans are slightly smaller than the corresponding 1P scans. This happens due to the non-linear response to 2P excitation, which laterally thins and axially shortens the excitation light-sheet. While excitation from a thinner light-sheet improves resolution, the concomitant axial shortening reduces maximal scanned volume. Lowering illumination NA of the beam would compensate for this reduction, but at a substantial cost in power, highlighting the compromise between axial extent and available excitation laser power in 2P SOPi. While 2P SOPi wins on resolution, 1P SOPi is capable of rapid volumetric imaging exceeding 10 volumes per second, limited by system NA, as well as camera sensitivity and speed. In the current SOPi implementation, we used an inexpensive laser diode to generate the 1P light-sheet. Since emission from a diode is divergent, we used a converging lens for collimation, followed by a slit aperture and a cylindrical lens for light-sheet generation. For many applications in relatively optically clear samples, including zebrafish, the inexpensive laser diode-based SOPi implementation is sufficient for 10 volumes per second (or higher) live imaging of neural activity across large structures in the brain and the spinal cord, with preserved steric access for concurrent electrophysiology or behavioral manipulations and no required post processing for 2D imaging.

In an alternative arrangement, it is possible to improve 1P SOPi imaging performance through use of a standard laser beam in a DSLM implementation [12], along with camera rolling shutter-assisted confocal slit detection [37]. Deconvolution approach [38] would further improve 1P and 2P SOPi imaging performance, but the latter implementation would perform better in optically scattering samples, such as mouse brains *in vivo*. Currently, SOPi has relatively small overall numerical aperture, but superior high-cost objectives could replace the current ones to increase the overall system NA, pushing the attainable resolution to sub-micron scale, potentially into the domain of single molecule imaging. Use of higher NA objectives would help gather more light, also increasing the overall imaging speed for both 1P and 2P implementations, since it is not currently limited by galvo speed. The current 2P light-sheet implementation is based on scanning low NA Gaussian beams. A Bessel beam 2P light-sheet implementation [39] would further improve the penetration depth and resolution of SOPi.

In its scanning-descanning optical arrangement, SOPi may appear similar to a confocal theta line-scanning microscope [40]. Key differences include low NA illumination, absence of a confocal slit in detection arm and, most importantly, SOPi's high sensitivity to placement and construction of the galvo scanner. While confocal theta line-scanning is usually performed with a polygon mirror based scanner, SOPi implementation mandates a plane mirror based galvo scanner placed in the conjugate plane to the back focal plane of the main objective. We have demonstrated that this is required for constant tilt light-sheet scanning. While no prior publications demonstrate 2P oblique plane light-sheet microscopy systems, a proof of concept 2P variant of SCAPE was recently presented [41]. The 2P SCAPE approach relies on scanning and descanning the laser line along the orthogonal (x and y) axis, to be imaged onto central rows of a sCMOS camera. This implementation is highly restricted in camera pixel use and would require stitching and post processing of single row pixel data to reconstruct 2D image sections and 3D volumes. The authors of 2P SCAPE did not appear to modify the scanning architecture that causes scan position-dependent variance in tilt. In comparison, SOPi's optimized scanning architecture and the use of all camera pixels supports live visualization of any sectioned plane within a sample. The minimal post processing in the form of affine transformation is only required for visualization of entire 3D volumes acquired by SOPi.

In summary, we have implemented and improved single objective based scanned oblique planar excitation microscopy. In comparison to other existing single objective based light-sheet microscopy approaches, our SOPi implementation is characterized by a simplified design and allows for true shape 3D reconstruction of scanned volume. The simple design of SOPi makes it easy to expand functionality in the system. Straightforward future modifications include extending the system for simultaneous, multichannel imaging by introduction of an emission splitting system. Single objective based light-sheet microscopy can also include lattice light-sheet [42] or Airy light-sheet [43] approaches with further modification in the illumination architecture. SOPi's simplified volume reconstruction can easily find use in biological imaging applications where shape-related quantitative volumetric measurements are important.

## Author contributions

MK and YK conceived the experiments. MK performed ray-tracing simulations, built the SOPi system, and performed imaging experiments. JN prepared fixed mouse brain slices. SK and DM prepared and imaged zebrafish samples. MK and YK wrote the manuscript with inputs from all authors.

## Acknowledgments

We thank Lindsey Butler for genotyping and mouse colony management, and Elissa Szuter for technical help maintaining the zebrafish colony.

## Funding

This work was supported by the Beckman Young Investigator Award, Searle Scholar Award, William and Bernice E. Bumpus Young Innovator Award, Rita Allen Foundation Scholar Award, and Sloan Research Fellowship (Y.K.), and by R01-NS067299 (D.L.M).

## Disclosure

A provisional patent has been filed based on this work.